\documentclass[letterpaper]{article}

\usepackage{emulateapj}
\usepackage{onecolfloat}
\usepackage{apjfonts}
\usepackage{amsmath}
\usepackage{natbib}
\usepackage{graphicx}

\newcommand{\Msun}      {\mbox{$\rm\,M_{\mathord\odot}$}}

\submitted{Accepted by the Astrophysical Journal}

\begin{document}

\twocolumn[
\title{X-Ray Jet Emission from the Black Hole X-Ray Binary 
XTE~J1550--564\\ with {\em Chandra} in 2000}

\author{John A. Tomsick}
\affil{Center for Astrophysics and Space Sciences, Code
0424, University of California at San Diego, La Jolla, CA,
92093, USA (e-mail: jtomsick@ucsd.edu)}

\vspace{0.1cm}
\author{St\'ephane Corbel}
\affil{Universit\'e Paris VII and Service d'Astrophysique, 
CEA Saclay, 91191 Gif sur Yvette, France}

\author{Rob Fender}
\affil{Astronomical Institute ``Anton Pannekoek,'' University
of Amsterdam and Center for High Energy Astrophysics, Kruislaan 403, 
NL-1098 SJ Amsterdam, The Netherlands}

\vspace{0.1cm}
\author{Jon M. Miller}
\affil{Department of Physics and Center for Space Research, 
Massachusetts Institute of Technology, Cambridge, MA 02139, USA}

\author{Jerome A. Orosz}
\affil{Astronomical Institute, Utrecht University, Postbus 80000,
3508 TA Utrecht, The Netherlands}

\author{Tasso Tzioumis}
\affil{ATNF, CSIRO, P.O. Box 76, Epping, NSW 1710, Australia}

\author{Rudy Wijnands}
\affil{Department of Physics and Center for Space Research, 
Massachusetts Institute of Technology, Cambridge, MA 02139, USA and Chandra Fellow}

\vspace{0.2cm}
\author{Philip Kaaret}
\affil{Harvard-Smithsonian Center for Astrophysics, 60 Garden Street,
 Cambridge, MA, 02138, USA}

\begin{abstract}

We have discovered an X-ray jet due to material ejected from the
black hole X-ray transient XTE~J1550--564.  The discovery was first 
reported by \cite{corbel02b}, and here, we present an analysis of 
the three {\em Chandra} observations made between 2000 June and 2000 
September.  For these observations, a source is present that moves
in an eastward direction away from the point source associated with 
the compact object.  The separation between the new source and the 
compact object changes from $21^{\prime\prime}.3$ in June to 
$23^{\prime\prime}.4$ in September, implying a proper motion of 
$21.2\pm 7.2$~mas~day$^{-1}$, a projected separation of 0.31-0.85~pc 
and an apparent jet velocity between $0.34\pm 0.12$ and $0.93\pm 0.32$ 
times the speed of light for a source distance range of 
$d = 2.8$-7.6~kpc.  These observations represent the first time that 
an X-ray jet proper motion measurement has been obtained for any 
accretion powered Galactic or extra-galactic source.  While this work 
deals with the jet to the east of the compact object, the western jet 
has also been detected in the X-ray and radio bands.  The most likely 
scenario is that the eastern jet is the approaching jet and that the 
jet material was ejected from the black hole in 1998.  Along with a 
1998 VLBI proper motion measurement, the {\em Chandra} proper motion 
indicates that the eastern jet decelerated between 1998 and 2000.  
There is evidence that the eastern jet is extended by 
$\pm 2^{\prime\prime}$-$3^{\prime\prime}$ in the direction of the 
proper motion.  The upper limit on the source extension in the 
perpendicular direction is $\pm 1^{\prime\prime}.5$, which 
corresponds to a jet opening angle of $<$$7.5^{\circ}$.  The X-ray 
jet energy spectrum is well- but not uniquely described by a power-law 
with an energy index of $\alpha = -0.8\pm 0.4$ ($S_{\nu}\propto\nu^{\alpha}$)
and interstellar absorption.  The eastern jet was also detected in the radio 
band during an observation made within 7.4~days of the June {\em Chandra} 
observation.  The overall radio flux level is consistent with an extrapolation 
of the X-ray power-law with $\alpha = -0.6$.  The 0.3-8~keV X-ray jet 
luminosity is in the range (3-24)$\times 10^{32}$~erg~s$^{-1}$ for the 
June observation using the distance range above but is a factor of 
$\sim$2-3 lower for the later observations.  We cannot definitively
determine the X-ray emission mechanism, but a synchrotron origin
is viable and may provide the simplest explanation for the observations.

\end{abstract}

\keywords{acceleration of particles --- accretion, accretion disks ---
black hole physics --- stars: individual (XTE~J1550--564) --- 
stars: winds, outflows --- X-rays: stars}

] 

\section{Introduction}

Outflows are observed in active galactic nuclei (AGN) and for some Galactic 
compact objects containing relativistic particles that are accelerated away 
from the compact objects in collimated jets.  At least three types of radio 
jets are observed in Galactic X-ray binaries.  In 1992, double-sided radio 
lobes were detected for two accreting black hole candidates:  GRS~1758--258 
\citep{rmm92} and 1E~1740.7--2942 \citep{mirabel92}.  For these sources, the 
compact object/radio lobe separations are 1-3~pc, and the lobes are stationary.  
Due to observational and likely physical similarities to AGN, the name 
microquasar was given to these sources.  The number of X-ray binaries in 
the group of microquasars was greatly increased with the discovery of 
relativistic radio jets on much smaller size scales (0.02-0.06~pc).  The 
two best known systems are GRS~1915+105 \citep{mr94} and GRO~J1655--40 
\citep{tingay95,hr95}.  For both of these systems, apparently superluminal jet 
velocities are observed, and the actual jet velocities inferred are $>$0.9c.  
Although not as relevant for this work, the third type of radio jet is often 
called a ``compact jet'' and has been detected for a relatively large number 
of X-ray binaries.  Despite their small size ($\sim$$10^{-4}$~pc), compact 
jets have been resolved for GRS~1915+105 \citep{dhawan00} and Cyg~X-1 
\citep{stirling01}.

The jets are usually detected at radio wavelengths, but, in AGN, optical 
and X-ray jets are also frequently seen.  With the exception of the 
large-scale ($\sim$40~pc) diffuse X-ray emission detected from the X-ray 
binary SS~433 with the {\em Einstein Observatory} \citep{seward80}, X-ray 
jets were not seen for Galactic systems prior to the launch of the 
{\em Chandra X-ray Observatory} \citep{weisskopf02} in 1999.  With a large 
improvement in angular resolution over previous missions, {\em Chandra} 
detected arcsecond ($\sim$0.025~pc) X-ray jets in SS~433 \citep{marshall02}, 
but similar jets have not previously been reported for other accretion
powered Galactic sources.  

XTE~J1550--564 was first detected by the {\em Rossi X-ray Timing 
Explorer (RXTE)} All-Sky Monitor (ASM) in 1998 September \citep{smith98}.  
It was identified as a probable black hole system based on its X-ray 
spectral and timing properties, and optical observations of the
source in quiescence indicate a compact object mass near 10\Msun, 
confirming that the system contains a black hole \citep{orosz02}.
Soon after the discovery of the source, a jet ejection with an 
apparent separation velocity $>$2c was observed in the radio band 
using Very Long Baseline Interferometry (VLBI), establishing 
that the source is a microquasar \citep{hannikainen01}.  The VLBI 
observations followed a very bright radio and X-ray flare that was 
likely related to the ejection event \citep{sobczak99,wu02}.  
XTE~J1550--564 has shown a high degree of X-ray activity over the last 
few years, having a major outburst in 2000 
(Tomsick, Corbel \& Kaaret 2001\nocite{tck01} and references therein)
and mini-outbursts in 2001 and 2002 \citep{tomsick01,ssm02}.  Bright radio 
and X-ray flares like the 1998 flare have not been observed during these 
outbursts, but unresolved radio emission was detected, indicating the 
presence of a compact jet \citep{corbel01}.

During the 2002 X-ray outburst, radio observations of XTE J1550--564
were made with the {\em Australia Telescope Compact Array (ATCA)}.
The detection of a variable radio source $22^{\prime\prime}$ west of 
XTE~J1550--564 \citep{corbel02a} prompted us to examine {\em Chandra} 
observations of the source that were made in 2000.  As reported by 
\cite{corbel02b}, we discovered an X-ray source that is at approximately 
the same angular distance from XTE~J1550--564 as the variable radio 
source and is to the east.  In this paper, we present an analysis of 
the {\em Chandra} observations.  The three sources (XTE~J1550--564, 
the western radio source and the eastern X-ray source) are aligned with 
each other and with the VLBI radio jets that were detected in 1998.  
The alignment strongly suggests that the X-ray and radio sources are 
jets from XTE~J1550--564.  The jet size scale that we infer is in 
between those of the previously detected radio jets, providing an 
important connection between these two types of jets.

\section{Observations}

Seven XTE~J1550--564 {\em Chandra} observations occurred in 
May and June during the 2000 outburst from this source, and 
two observations were made near the end of the outburst on 
2000 August 21 and September 11.  These observations were 
described in previous work where results for the XTE~J1550--564 
point source were presented \citep{tck01,miller01}.  Modes with 
two-dimensional imaging were used for the observations made on 
June 9, August 21 and September 11 (henceforth observations 1, 
2 and 3), and, here, we focus on these observations.  The rest 
of the May and June observations were grating observations made 
in ``continuous clocking mode,'' providing only one-dimensional 
imaging.  In addition, a filter was used to block the zeroth
order due to the high XTE~J1550--564 flux level.  Thus, with
the exception of the June 9 observation, the May and June
observations are not suitable for our search for extended 
emission from the source.  

Table~\ref{tab:obs} provides information about observations
1, 2 and 3, including the observation time, the exposure 
time and the instrument configuration.  All three 
observations were made using the Advanced CCD Imaging
Spectrometer (ACIS).  In each case, XTE~J1550--564 was 
placed on one of the back-illuminated ACIS chips (S3), 
providing the best low-energy response.  Observation 1 
differs from the other two observations in that the
High Energy Transmission Grating (HETG) was inserted.
Although grating observations provide two-dimensional
imaging, the sensitivity is reduced.  Also, observation 1 
includes data taken with two different ACIS CCD 
configurations with longer (1.1~s) and shorter (0.3~s) 
individual frame exposure times (``alternating exposure
mode'').  We carried out our analysis using the data from 
all exposures with a total exposure time of 4957~s.
We made background light curves for all three observations 
and found that, for observation 2, there are brief time 
segments where excess background is observed.  Removing
these causes a drop in the exposure time from 5099 to
4985~s.  Periods of high background are not seen for
the other observations.  For observation 3, we used
data from the entire 4572~s exposure time.  

\begin{table}[t]
\caption{XTE~J1550--564 {\em Chandra} Observations\label{tab:obs}}
\begin{minipage}{\linewidth}
\footnotesize
\begin{tabular}{c|c|c|c|c} \hline \hline
 & MJD\footnote{Modified Julian Date (JD$-$2400000.5) at exposure midpoint.} & 
 & Exposure & Instrument\\
Observation & (days) & UT Date & Time (s) & Configuration\\ \hline
1 & 51704.538 & 2000 June 9 & 4957 & ACIS-S/HETG\\
2 & 51777.405 & 2000 August 21 & 4985 & ACIS-S\\
3 & 51798.245 & 2000 September 11 & 4572 & ACIS-S\\
\end{tabular}
\end{minipage}
\end{table}

While a detailed analysis of the XTE~J1550--564 energy 
spectrum is presented for observations 2 and 3 in 
\cite{tck01}, similar results for observation 1 have 
not been reported.  Thus, we extracted the ACIS grating 
spectrum for observation 1.  We fitted 0.8-8 keV MEG 
(Medium Energy Grating) and 2-10 keV HEG (High Energy 
Grating) energy spectra with a model consisting of a 
power-law with interstellar absorption.  A reasonably 
good (but not formally acceptable) fit is obtained 
($\chi^{2}/\nu = 94/71$) with a column density of 
$N_{\rm H} = (8.4\pm 0.8)\times 10^{21}$~cm$^{-2}$, 
which is consistent with the Galactic value in the 
direction of XTE~J1550--564, and a power-law photon index
of $\Gamma = 1.58\pm 0.08$ (90\% confidence errors).  
The fit is not significantly improved by the addition
of a soft ``disk-blackbody'' component \citep{makishima86},
and no obvious emission lines are present in the residuals.
The 0.3-8~keV source flux for observation 1 is 
$9.3\times 10^{-11}$ erg~cm$^{-2}$~s$^{-1}$.  For 
observations 2 and 3, we obtain 0.3-8~keV source fluxes 
of $1.2\times 10^{-13}$ and 
$2.4\times 10^{-13}$ erg~cm$^{-2}$~s$^{-1}$, respectively, 
using the spectral parameters from \cite{tck01}.

This paper also includes a re-analysis of a 5~hr radio 
band observation made with {\em ATCA} on 2000 June 1, 
which is close in time to {\em Chandra} observation 1.  
The {\em ATCA} observation was carried out in the high 
spatial resolution 6B array configuration, and flux
measurements were obtained at central frequencies of 
1384, 2496, 4800 and 8640~MHz with a total bandwidth 
of 128 MHz.  More details can be found in a previous 
publication where these observations were used \citep{corbel01}.

\section{Results}

We searched for X-ray sources in the 0.3-8~keV images for all 
three {\em Chandra} observations using the ``wavdetect'' program 
\citep{freeman02} in version 2.2 of the CIAO ({\em Chandra} 
Interactive Analysis of Observations) software package.  We only 
included photons with energies up to 8~keV as the ACIS effective 
area drops off rapidly above this energy.  We used data from the 
ACIS chip containing XTE~J1550--564 (ACIS-S3).  For observation 1, 
we restricted the search to a square region of dimension 
1$^{\prime}$.5 centered on XTE~J1550--564 to avoid regions 
that include the dispersed spectrum, while we searched 
over the entire S3 chip for observations 2 and 3.  

For observation 1, only one other source besides XTE~J1550--564
is detected at a significance level of 4.8-$\sigma$.  The 
source is $\sim$21$^{\prime\prime}$ east of XTE~J1550--564, 
while the new radio transient is $\sim$22$^{\prime\prime}$ 
west of XTE~J1550--564.  For observation 2, there are seven
sources detected at $>$3-$\sigma$ significance, excluding 
XTE~J1550--564.  Although none of the sources have positions 
consistent with that of the new radio transient (see Kaaret 
et al.~2002\nocite{kaaret02} for flux limits), the most 
significantly detected source (at 8.3-$\sigma$) is located 
$\sim$23$^{\prime\prime}$ east of XTE~J1550--564.  For 
observation 3, there are two sources detected at $>$3-$\sigma$ 
significance, excluding XTE~J1550--564.  Similarly to 
observation 2, the most significantly detected source 
(at 9.1-$\sigma$) is located $\sim$23$^{\prime\prime}$ east 
of XTE~J1550--564, but we do not detect a source at the 
position of the new radio transient.

\begin{figure}[t]
\plotone{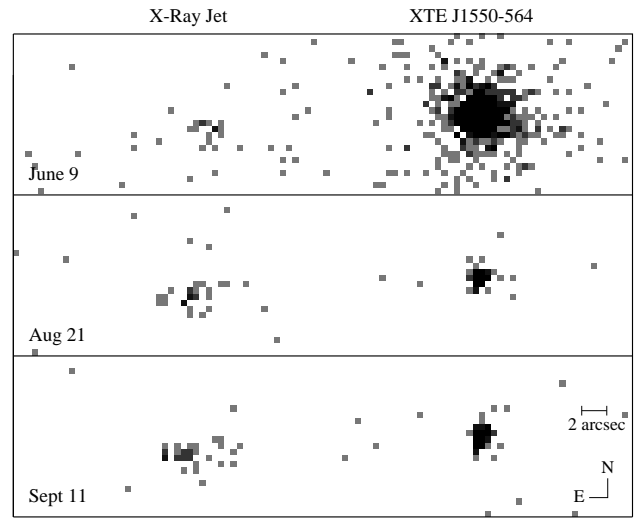}
\vspace{0.3cm}
\caption{Three {\em Chandra} 0.3-8~keV images showing XTE~J1550--564 
and an X-ray jet.  The observations were made in 2000 on June 9, August 21 
and September 11, and exposure times of 4957~s, 4985~s and 4572~s were obtained, 
respectively.  The greyscale is logarithmic based on the number of counts
detected per pixel.  To provide a scale where XTE~J1550--564 and the X-ray 
jet are both visible, we set a saturation level of 12 counts.  We used
the same scale in all three images, but it should be noted that the levels 
for June 9 are not directly comparable to those for the other two 
observations since a grating was inserted for the June 9 observation.
The pixel size is $0^{\prime\prime}.492$-by-$0^{\prime\prime}.492$.
\label{fig:rimage}}
\end{figure}

Figure~\ref{fig:rimage} shows the portion of the 0.3-8~keV raw 
(i.e., un-smoothed) images for the three observations containing 
XTE~J1550--564 and the source to the east of XTE~J1550--564 that 
is detected in all three observations.  For observation 3, the 
position of the eastern source is R.A. = 15h 51m 01s.47, 
Dec. = --56$^{\circ}$~28$^{\prime}$~36$^{\prime\prime}$.7 
(equinox 2000.0).  We calculated this position using the 
XTE~J1550--564 radio position, which is known to 
$0^{\prime\prime}.3$ \citep{corbel01}, and the separations 
between XTE~J1550--564 and the X-ray jet in the {\em Chandra}
images.  The separation measurement is good to about 
$0^{\prime\prime}.5$ (see \S3.2), indicating an error of 
less than $0^{\prime\prime}.8$ on the X-ray jet position.  
While we do not detect an X-ray source in these observations 
coincident with the western radio source, it is notable that 
the three sources (XTE~J1550--564, the western radio source 
and the eastern X-ray source) are aligned 
(see Corbel et al.~2002b\nocite{corbel02b} for a clear 
illustration of the alignment).  In addition,  it is 
remarkable that they are aligned with the radio jets that
were resolved during the bright 1998 X-ray outburst from 
XTE~J1550--564 \citep{hannikainen01}.  This provides evidence 
for a connection between the radio jets, the eastern X-ray 
source and the western radio source, and the results presented 
below support this interpretation.  Henceforth, we refer to 
the eastern X-ray source as the X-ray jet.  

\subsection{Proper Motion of the X-Ray Jet}

We determined the separation between XTE~J1550--564 and the
X-ray jet for each observation.  While an inspection of the 
images indicates that we obtain reasonable X-ray jet positions 
with wavdetect, we used a second measurement technique to 
obtain an estimate of the systematic error.  For each observation, 
we determined the positions for XTE~J1550--564 and the X-ray jet 
by calculating the source centroid using the 0.3-8~keV events 
from a 16-by-16 pixel ($7^{\prime\prime}.9$-by-$7^{\prime\prime}$.9) 
region centered on the wavdetect position.  An inspection of the 
images (see Figure~\ref{fig:rimage}) indicates that this region 
contains all or nearly all of the events that are likely related 
to the source along with a small number of background events.
We estimate that the number of background events in the 16-by-16 
pixel regions are 4.0, 1.2 and 0.7 for observations 1, 2 and 3, 
respectively.  Our technique for estimating the background surface 
brightness is described in \S3.3.  Compared to the wavdetect 
positions, the re-calculated XTE~J1550--564 positions change by 
less than $0^{\prime\prime}.1$, and the X-ray jet positions change 
by less than $0^{\prime\prime}.5$.  For observations 1 and 2, the 
re-calculated source separations are $0^{\prime\prime}.45$ and 
$0^{\prime\prime}.27$ larger, respectively, and, for observation 3, 
the re-calculated source separation is $0^{\prime\prime}.38$ smaller.  
Based on these changes, we estimate that the error on the source 
separation is approximately $0^{\prime\prime}.5$ and is dominated
by systematics.

\begin{table}[b]
\caption{X-Ray Jet Results\label{tab:results}}
\begin{minipage}{\linewidth}
\footnotesize
\begin{tabular}{c|c|c|c|c} \hline \hline
 & Rate\footnote{0.3-8~keV count rate after background subtraction.} & 
Separation\footnote{Separation between the XTE~J1550--564 point source and the 
X-ray jet.} &  & \\
Observation & (counts~s$^{-1}$) & (arcseconds) & 
$\Gamma$\footnote{Power-law photon index (with 90\% confidence errors) obtained 
with the column density fixed to the Galactic value of 
$N_{\rm H} = 9\times 10^{21}$~cm$^{-2}$} & Flux\footnote{Absorbed 0.3-8~keV X-ray 
flux in erg~cm$^{-2}$~s$^{-1}$ obtained from a simultaneous power-law fit to all 
three observations.  We fixed $N_{\rm H}$ to the Galactic value and found a photon 
index of $\Gamma = 1.8\pm 0.4$.}\\ \hline
1 & $3.0\times 10^{-3}$ & $21.3\pm 0.5$ & $1.7^{+0.7}_{-1.0}$ & $(20\pm 6)\times 10^{-14}$\\
2 & $5.0\times 10^{-3}$ & $22.7\pm 0.5$ & $1.8\pm 0.6$ & $(6.1\pm 1.3)\times 10^{-14}$\\
3 & $6.9\times 10^{-3}$ & $23.4\pm 0.5$ & $1.9\pm 0.8$ & $(8.2\pm 1.5)\times 10^{-14}$\\
\end{tabular}
\end{minipage}
\end{table}

\begin{figure}[b]
\plotone{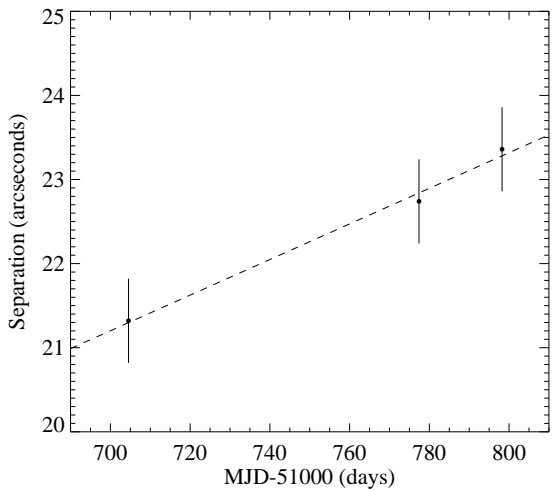}
\caption{Angular separation between XTE~J1550--564 and the X-ray
jet for the June 9, August 21 and September 11 {\em Chandra}
observations.  The dashed line shows a linear fit corresponding
to a proper motion of $21.2\pm 7.2$~mas~day$^{-1}$, and an 
extrapolation of the linear fit predicts zero separation on 
MJD~$50699\pm 278$.
\label{fig:pm}}
\end{figure}

The angular separations between XTE~J1550--564 and the X-ray jet
are given in Table~\ref{tab:results} and are plotted vs.~time in 
Figure~\ref{fig:pm}.  The separations show that the X-ray jet is 
moving away from XTE~J1550--564, and the data are well-described
by a linear increase in the separation.  The linear fit shown in 
the figure corresponds to a proper motion of 
$21.2\pm 7.2$~mas~day$^{-1}$ (mas = milliarcseconds), and an 
extrapolation of the linear fit predicts zero separation on 
MJD~$50699\pm 278$.  MJD~50699 corresponds to 1997 September 8, 
which is nearly a year before the first reported X-ray and radio 
activity for XTE~J1550--564 in 1998 September.  We obtained the 
{\em RXTE}/ASM light curve for the full {\em RXTE} mission to 
check for X-ray activity between early 1996 and 1998 September.  
XTE~J1550--564 was not detected during this time with a 3-$\sigma$ 
upper limit of $1\times 10^{-9}$~erg~cm$^{-2}$~s$^{-1}$ for daily 
1.5-12~keV flux measurements.  The largest gap in ASM coverage 
during this time was 14~days, which is much shorter than the 
duration of a typical outburst.  Thus, it is most likely that the 
material responsible for the X-ray jet was ejected during the 
large 1998 September radio and X-ray flare \citep{wu02}.  We 
note that the errors on the linear fit parameters are dominated 
by the systematic errors on the individual separations and, thus, 
do not correspond to 68\% confidence statistical errors.

\subsection{Source Morphology}

\begin{figure}[t]
\plotone{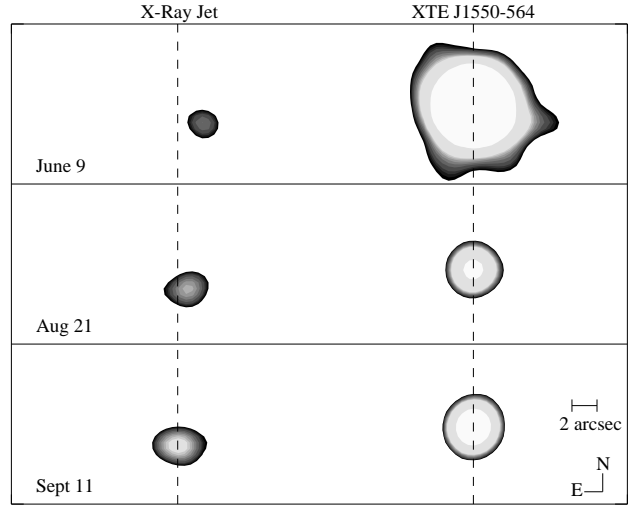}
\caption{Contour plots produced by convolving the 0.3-8~keV
images shown in Figure~\ref{fig:rimage} with a 2-dimensional 
Gaussian with a width ($\sigma$) of 2 pixels in both directions.  
The same contour levels are used for all three images.  The 
vertical dashed lines clearly demonstrate the X-ray jet motion 
relative to XTE~J1550--564.  In addition, the August 21 and 
September 11 images suggest that the X-ray jet is elongated 
in the direction of the proper motion.
\label{fig:image}}
\end{figure}

We produced the contour plots shown in Figure~\ref{fig:image}
for the three observations after convolving the raw images (from 
Figure~\ref{fig:rimage}) with a 2-dimensional Gaussian with a 
width ($\sigma$) of 2 pixels in both directions.  In addition 
to illustrating that the X-ray jet moves away from XTE~J1550--564 
over time, the contour plots for observations 2 and 3 suggest that 
the X-ray jet is elongated in the direction of the proper motion 
(roughly east-to-west).  

\begin{figure}[t]
\plotone{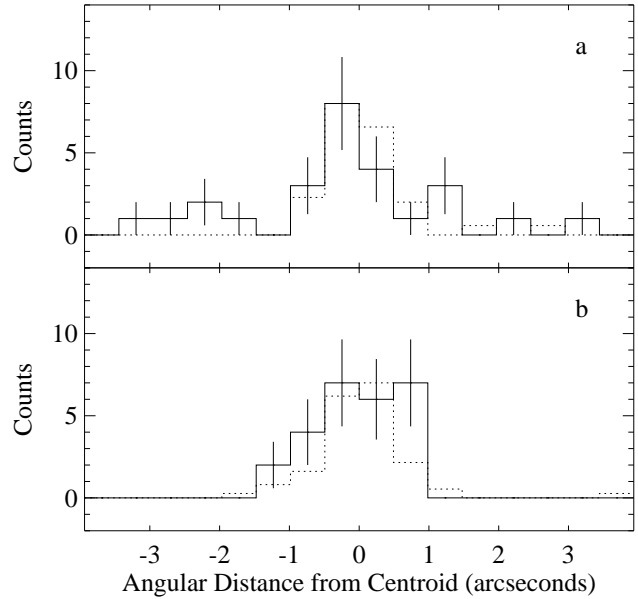}
\vspace{0.3cm}
\caption{Spatial profiles showing the number of counts as a 
function of position along the jet axis (a) and perpendicular
to the jet axis (b) for the observation 2 X-ray jet.  In both 
panels, the solid lines with Poisson error bars are the jet 
profiles and the dotted lines are the XTE~J1550--564 profiles
after re-normalizing to facilitate comparison to the X-ray jet 
profiles.\label{fig:profiles1845}}
\end{figure}

\begin{figure}[t]
\plotone{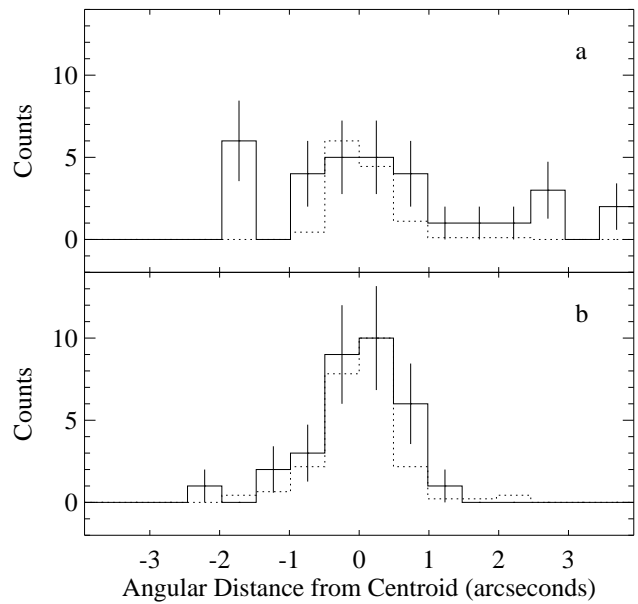}
\vspace{0.3cm}
\caption{Spatial profiles showing the number of counts as a 
function of position along the jet axis (a) and perpendicular
to the jet axis (b) for the observation 3 X-ray jet.  The line 
designations are the same as for Figure~\ref{fig:profiles1845}.
\label{fig:profiles1846}}
\end{figure}

To study the jet morphology for observations 2 and 3, we produced 
source profiles showing the number of counts as a function of 
position along and perpendicular to the jet axis.  We determined 
the jet axis using the source positions obtained above.  For 
observations 2 and 3, the jet position angles are 
$93.9^{\circ} \pm 1.3^{\circ}$ and $93.7^{\circ} \pm 1.3^{\circ}$, 
respectively.  Here, we use a value of $93.8^{\circ}$, which is 
consistent with the position angle of $93.9^{\circ} \pm 0.8^{\circ}$
derived for the VLBI radio jets (D. Hannikainen, private communication).  
In producing the source profiles, we used the source counts from a 
16-by-16 pixel ($7^{\prime\prime}.9$-by-$7^{\prime\prime}$.9) 
region centered on the jet position and oriented so that the 
re-sampled pixels run along and perpendicular to the jet axis.  
We then binned the counts in 1 pixel strips in both directions.  
Each strip is expected to contain 0.07 and 0.05 background counts 
for observations 2 and 3, respectively, and we neglected this low 
background level for this analysis.  We repeated the analysis 
for XTE~J1550--564 to provide a comparison to the jet profiles.
Figures \ref{fig:profiles1845} and \ref{fig:profiles1846} show 
the results for observations 2 and 3, respectively.  In each case, 
panels a and b are the profiles along and perpendicular to the
jet axis, respectively.  The solid lines with Poisson error bars 
are the jet profiles and the dotted lines are the XTE~J1550--564
profiles after re-normalizing so that the peak bin has the same 
number of counts for the jet and XTE~J1550--564.  

We performed Kolmogorov-Smirnov (KS) tests \citep{press92} on the 
un-binned counts to obtain a formal comparison of the profiles.
Using the KS statistic, we calculated the probability that the 
X-ray jet profiles along and perpendicular to the jet axis are 
taken from the same parent distribution as the XTE~J1550--564 
profiles (i.e., the probability that the X-ray jet is not extended 
in a direction).  In addition, we compared the X-ray jet profiles 
to the profiles for the source PG~1634+706 (Observation I.D. 1269), 
which is used to calibrate the ACIS point-spread-function (PSF).  
The results shown in Table~\ref{tab:profiles} indicate that it 
is unlikely that the X-ray jet profile for observation 3 is the 
same as the XTE~J1550--564 and PG~1634+706 profiles along the 
jet axis with probabilities of 1.8\% and 0.2\% for the respective 
comparison sources.  The extension along the jet axis is not
significant for observation 2 with probabilities of 21\% and 6.3\%
for XTE~J1550--564 and PG~1634+706, respectively.  In the direction 
perpendicular to the jet axis, the probabilities that the X-ray 
jet profiles are the same as the comparison source profiles are 
in the range 21-71\%.  Thus, there is no evidence that the X-ray 
jet is extended in this direction.  

\begin{table}[b]
\caption{KS Test Results for Comparing 1-Dimensional Profiles\label{tab:profiles}}
\begin{minipage}{\linewidth}
\footnotesize
\begin{tabular}{c|c|c|c} \hline \hline
 & & Probability & Probability\\
 & Comparison & (Along & (Perpendicular\\
Observation & Source\footnote{The X-ray jet profiles along the jet axis
and perpendicular to the jet axis are compared to the 1-dimensional 
profiles for the comparison sources.} & Jet Axis)\footnote{KS test probability
that the X-ray jet profile along the jet axis is the same as the comparison
profile.} & to Jet Axis)\footnote{KS test probability that the X-ray jet profile 
perpendicular to the jet axis is the same as the comparison profile.}\\ \hline
2 & XTE~J1550--564 & 0.208 & 0.465\\
2 & PG~1634+706 & 0.063 & 0.207\\
3 & XTE~J1550--564 & 0.018 & 0.707\\
3 & PG~1634+706 & 0.0024 & 0.378\\
\end{tabular}
\end{minipage}
\end{table}

In summary, the KS tests show that it is likely (at least for 
observation 3) that the X-ray jet is extended along the jet
axis.  From Figures \ref{fig:profiles1845}a and 
\ref{fig:profiles1846}a, we estimate that the jets are extended 
by $\pm 2^{\prime\prime}$-$3^{\prime\prime}$ along the jet axis, 
but by less than $\pm 1^{\prime\prime}.5$ in the perpendicular 
direction.  At an angular separation of $23^{\prime\prime}$ from 
XTE~J1550--564, the upper limit of $\pm 1^{\prime\prime}.5$ 
corresponds to a jet opening angle of $<$$7.5^{\circ}$.  

\subsection{X-Ray Jet Energy Spectrum and Flux}

We extracted energy spectra for each observation.  For 
observations 2 and 3, we used a circular source extraction
region with a radius of $4^{\prime\prime}$, and we extracted 
background spectra from an annulus with an inner radius of 
$5^{\prime\prime}$ and an outer radius of $18^{\prime\prime}$.
The source and background regions were both centered on the
X-ray jet.  For observation 1, the background count rate 
near the X-ray jet is several times higher than the rates for 
the other observations.  It is likely that this is partly due 
to the XTE~J1550--564 dust-scattering halo, but measurements of 
the observation 1 background rate far from XTE~J1550--564 show count 
rates that are still about a factor of two higher than the other 
observations, indicating that the intrinsic observation 1 background 
rate is somewhat higher.  Thus, we used a smaller circular source 
extraction region with a radius of $2^{\prime\prime}.5$ for 
observation 1 to minimize contamination.  Also, as the XTE~J1550--564 
X-ray halo contributes to the background near the X-ray jet, we 
estimated the background level using the counts in an annulus centered 
on XTE~J1550--564 with an inner radius of $18^{\prime\prime}.8$ and an 
outer radius of $23^{\prime\prime}.8$ ($\pm 2^{\prime\prime}.5$ from 
the angular separation between XTE~J1550--564 and the X-ray jet).  We 
removed the parts of the annulus containing the X-ray jet and the 
read-out strip.  In the 0.3-8~keV energy band, 17, 26 and 32 counts 
were collected in the source regions for observations 1, 2 and 3, 
respectively.  In the same energy band, we estimate background 
levels of 1.3, 1.0 and 0.6 counts in the source extraction regions
for observations 1, 2 and 3, respectively.  We used the CIAO routine 
``psextract'' to extract the spectra and create response matrices.

Although the low numbers of source counts do not provide high
quality spectra, we fitted the spectra in an attempt to
distinguish between models.  As the same instrumental configuration 
was used for observations 2 and 3, we combined the data from 
these two observations to improve the statistics.  For these 
initial fits, we re-binned the spectrum after combining the 
data from the two observations and fitted the spectra using 
$\chi^{2}$-minimization.  We used a power-law model, appropriate 
for non-thermal emission, and also a thermal bremsstrahlung model
(``bremss'' in XSPEC version 11.1) and included interstellar
absorption.  These two models provide fits of nearly identical 
quality, and in both cases the reduced-$\chi^{2}$ value is 
0.9 for 7 degrees of freedom.  The parameter constraints are
poor with both models giving a column density ($N_{\rm H}$)
range of approximately (2-14)$\times 10^{21}$~cm$^{-2}$
(90\% confidence, $\Delta\chi^2 = 2.7$).  With $N_{\rm H}$
left as a free parameter, the range of possible values for 
the power-law photon index is $\Gamma = 0.6$-2.5, and the
bremsstrahlung temperature is only constrained to be $>$2.8~keV.  
The column density range quoted above is consistent with the 
Galactic value in the direction of XTE~J1550--564 
($N_{\rm H} = 9\times 10^{21}$~cm$^{-2}$); thus, we fixed $N_{\rm H}$ 
to this value for the fits described below.  Detecting iron line 
emission would provide evidence that the X-ray emission has a thermal 
origin.  No clear excess is observed in the X-ray jet spectrum at 
iron line energies, but the upper limit on the equivalent width of 
such a line is high (several keV).  

We fitted the 0.3-8~keV spectra for each observation separately
with a power-law model and interstellar absorption with the column 
density fixed to the Galactic value.  Rather than $\chi^{2}$-minimization, 
we used the Cash statistic \citep{cash79}, which does not require that 
the data be re-binned in energy.  This is desirable since re-binning 
data necessarily removes spectral information.  We did not perform 
background subtraction as the background levels are low enough to 
neglect.  As shown in Table~\ref{tab:results}, we obtain 
$1.7^{+0.7}_{-1.0}$, $1.8\pm 0.6$ and $1.9\pm 0.6$ (with 90\%
confidence errors) for the power-law photon index ($\Gamma$) for 
observations 1, 2 and 3, respectively.  We conclude that there is no 
evidence for spectral variability.  Thus, we re-fitted the three 
spectra simultaneously and obtained a value of $\Gamma = 1.8\pm 0.4$
(energy index = $\alpha = -0.8\pm 0.4$, $S_{\nu}\propto\nu^{\alpha}$).  
Although we take this as our best estimate of the power-law photon 
index, we note that it was obtained with $N_{\rm H}$ fixed to the 
Galactic value and that uncertainties in $N_{\rm H}$ represent a 
possible source of systematic error.

We used the simultaneous power-law fit to the three spectra to
obtain measurements of the X-ray jet flux.  We obtain 0.3-8~keV
absorbed fluxes of $(20\pm 6)\times 10^{-14}$, 
$(6.1\pm 1.3)\times 10^{-14}$ and 
$(8.2\pm 1.5)\times 10^{-14}$~erg~cm$^{-2}$~s$^{-1}$
for observations 1, 2 and 3, respectively, indicating
that the flux dropped by a factor of $\sim$2-3 between the 
June observation and the August and September observations. 
The quoted errors are based on the numbers of source and 
background counts and Poisson statistics, and the observation 1 
flux is corrected for the reduction of sensitivity due to 
insertion of the grating.  A source of systematic error (not 
included in the quoted errors) comes from the fact that a smaller 
source extraction region was used for observation 1.  For 
observation 2 and 3, 15\% and 19\% of the counts within the 
source regions lie outside a radius of $2^{\prime\prime}.5$ 
(the observation 1 extraction radius), indicating that the 
observation 1 flux could be higher than the quoted value by 
15-20\%.  Even if this source of systematic error is considered, 
the data are not strongly inconsistent with a smooth exponential 
or power-law decay (see also Kaaret et al.~2002\nocite{kaaret02}).  
Based on the XTE~J1550--564 source distance 
($d = 2.8$-7.6~kpc, Orosz et al.~2002\nocite{orosz02}), the 
intrinsic 0.3-8~keV X-ray jet luminosity for observation 1 is in 
the range (3-24)$\times 10^{32}$~erg~s$^{-1}$.

\begin{figure}[t]
\plotone{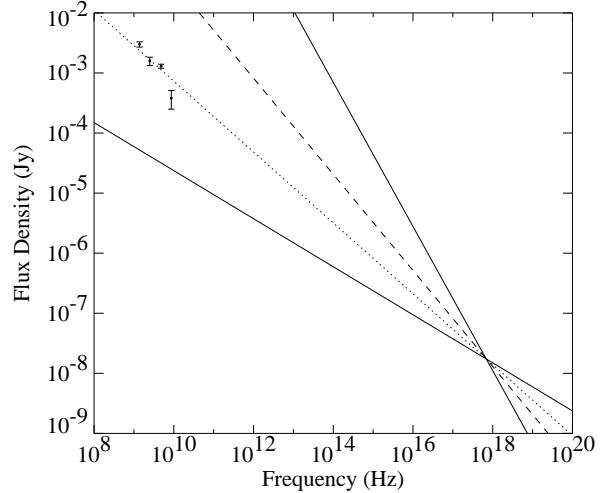}
\caption{Energy spectrum showing the power-law fits to the 
{\em Chandra} data (all 4 lines) for different assumed values 
of $\alpha$ and the radio flux measurements made on 2000 June 1 
with {\em ATCA}.  We use the normalization appropriate for 
observation 1 for the power-law fits as it occurred within 7.4~days 
of the radio observation.  The solid lines correspond to 
$\alpha = -0.4$ and $-1.2$ (90\% confidence lower and upper limits 
from the {\em Chandra} fits).  The dashed and dotted lines
correspond to $\alpha = -0.8$ and $-0.6$, respectively.
\label{fig:broadband}}
\end{figure}

We re-analyzed the 2000 June 1 {\em ATCA} radio band observations
and found a radio source coincident with the eastern jet.
We measured flux levels of $3.0\pm 0.3$, $1.58\pm 0.25$, 
$1.30\pm 0.10$ and $0.38\pm 0.13$~mJy at frequencies of
1384, 2496, 4800 and 8640~MHz.  We fitted the radio spectrum
with a power-law model and obtained an energy index of
$\alpha = -0.82\pm 0.15$ and a flux density of 
$1.1\pm 0.1$~mJy at 5000~MHz (90\% confidence errors in both 
cases).  The power-law fit is not formally acceptable with
a reduced-$\chi^{2}$ near 5 for 2 degrees of freedom, but
the overall radio spectrum is consistent with being an optically 
thin synchrotron spectrum in the sense that $\alpha < 0$.

Figure~\ref{fig:broadband} shows the four radio flux measurements 
along with the power-law fits to the {\em Chandra} energy spectra.  
We obtained the power-law fits to the {\em Chandra} data by fixing 
the power-law index to various values and fitting for the normalization.  
The normalization appropriate for observation 1 is used as
it occurred closest in time to the radio observations.  The 
power-law fits shown in the figure include the 90\% confidence
lower and upper limits on $\alpha$ ($-0.4$ and $-1.2$), the best 
fit value from the {\em Chandra} data ($-0.8$) and the value that 
is most consistent with the radio data ($-0.6$).  The ``pivot 
point'' for the power-law fits occurs at 2.9~keV, and the flux 
density is 17~nJy at that point.  The overall radio flux level 
is consistent with an extrapolation of the X-ray spectrum
with $\alpha = -0.6$, and it is possible that the radio and
X-ray emission are part of the same spectral component, which
would provide evidence for a synchrotron X-ray emission mechanism.  
We note that $\alpha = -0.6$ lies slightly outside the error region 
for $\alpha$ found from fitting the radio spectrum alone.  Although 
the X-ray and radio observations were not strictly simultaneous, 
observation 1 and the radio observation were only separated by 
7.4~days, and it is unlikely that the flux levels could change 
significantly over this time.

We also fitted the X-ray spectra for the three {\em Chandra}
observations simultaneously using the thermal bremsstrahlung 
model.  We used the Cash statistic and fixed the column density
to $9\times 10^{21}$~cm$^{-2}$.  The 90\% confidence lower
limit on the bremsstrahlung temperature is 3.8~keV, and the 
only other free parameter in the fit is the normalization, 
which is related to physical parameters according to
$N = 2.4\times 10^{-16} d^{-2} \int{n_{e} n_{i} {\rm d}V}$ cm$^{-5}$
\citep{rl79,arnaud96}, where $d$ is the source distance, and
$n_{e}$ and $n_{i}$ are, respectively, the electron and ion number 
densities within the volume $V$.  From the simultaneous fit, 
$N$ = (4-10)$\times 10^{-5}$~cm$^{-5}$ (90\% confidence)
for observation 1.

\section{Discussion}

These results demonstrate that the eastern X-ray source detected 
in the 2000 {\em Chandra} observations is, in fact, X-ray jet 
emission from XTE~J1550--564.  In addition to the alignment between 
the XTE~J1550--564 point source, the western radio source and the 
eastern X-ray source and the fact that these sources are aligned 
with the 1998 radio jets \citep{corbel02b}, the {\em Chandra}
images indicate motion of the eastern X-ray source away from 
XTE~J1550--564 along the jet axis and there is evidence that the 
eastern X-ray source is elongated in the direction of this motion.  
This is only the second Galactic X-ray binary, after SS~433 
\citep{seward80,marshall02}, where X-ray jet emission has been 
resolved, and these observations represent the first time that 
an X-ray jet proper motion measurement has been obtained for any 
accretion powered Galactic or extra-galactic source.

\subsection{Jet Kinematics}

Our proper motion measurement of $21.2\pm 7.2$~mas~day$^{-1}$ for 
the eastern X-ray jet and the source distance range of 2.8-7.6~kpc,
determined from optical observations of XTE~J1550--564 in X-ray
quiescence \citep{orosz02}, indicate an apparent jet velocity 
($\beta_{app} = v_{app}/c$) between $0.34\pm 0.12$ and $0.93\pm 0.32$.  
To make progress in estimating the actual jet velocity ($\beta$), 
it is necessary to determine what constraints we can place on the 
source geometry.  As jet emission is detected on both sides of 
XTE~J1550--564, we adopt the usual picture of a bipolar outflow 
with one approaching jet and one receding jet.  Although perfect 
symmetry between the two jets is not critical to the discussion 
we present here, we note that the fact that one of the jets was 
detected first in X-rays and the other was detected first in the 
radio should not be taken as evidence for a lack of symmetry.
In fact, another {\em Chandra} observation made in 2002 March 
indicates that there is X-ray emission coincident with the 
western jet \citep{kaaret02}.  

The main evidence that the eastern jet is the approaching jet is 
that the X-ray and radio observations made in 2000 and 2002 
indicate that the mean proper motion relative to XTE~J1550--564 
is greater for the eastern jet.  We determine the mean proper 
motions assuming that the material for both jets was ejected 
during the large 1998 September radio and X-ray flare 
\citep{wu02,sobczak99}.  \cite{corbel02b} present arguments for
this interpretation.  In addition, as mentioned above, an 
extrapolation of the linear fit to the X-ray jet/XTE~J1550--564 
separations predicts zero separation prior to 1998 September.  
Since the first detected activity from XTE~J1550--564 occurred 
in 1998 September, it is very unlikely that the ejection time 
was earlier.  We assume an ejection time of MJD 51078, which 
corresponds to the peak of the 1998 September radio flare 
\citep{wu02} and is 2~days before the resolved radio jets 
were detected \citep{hannikainen01}.  For the western jet, 
the separation between the jet and XTE~J1550--564 was 
$22^{\prime\prime}.43$ on MJD 52307 \citep{corbel02b}, implying 
a mean proper motion of 18.3~mas~day$^{-1}$.  Based on the 
linear fit to the three X-ray jet measurements, the X-ray jet 
reached $22^{\prime\prime}.43$ on MJD 51759, which implies a mean 
proper motion of 32.9~mas~day$^{-1}$.  The larger mean proper 
motion for the eastern jet indicates that it is the approaching 
jet.

Although there is strong evidence that the eastern jet is the 
approaching jet, the actual jet velocity ($\beta$) depends 
on the exact angle the jet axis makes with the line-of-sight 
($\theta$), which cannot be determined from the jet observations 
presented in this work.  Here, we derive a lower limit on the 
value of $\beta$ during 2000 June-September when the X-ray jet 
proper motion measurement was made.  The lower limit is independent 
of $\theta$ but depends on the distance range given above.  In general, 
for an approaching jet, the minimum value of $\beta$ occurs when 
$\theta_{min} = \tan^{-1}(1/\beta_{app})$.  Using the minimum 
apparent velocity for the X-ray jet ($\beta_{app} = 0.34\pm 0.12$)
and the equation
$\beta = \beta_{app}/(\sin\theta_{min} + \beta_{app}\cos\theta_{min})$, 
which relates the apparent velocity to the actual velocity for
an approaching jet, we obtain a minimum value for $\beta$
of $0.32\pm 0.10$ for $\theta_{min} = 71^{\circ}$.  Thus, we conclude 
that the X-ray jet is moving at a velocity of at least 22\% of the 
speed of light.  

\subsection{X-Ray Emission Mechanism}

For the X-ray jets in AGN, three emission mechanisms are commonly 
considered:  thermal bremsstrahlung, synchrotron and inverse Compton 
\citep{hk02}.  For most AGN jets, observations rule out thermal 
bremsstrahlung.  When high quality X-ray jet spectra have been 
obtained for AGN (e.g., Marshall et al.~2001\nocite{marshall01}), 
thermal emission lines have not been detected.  Further, the very 
large amount of hot gas necessary to explain the emission via thermal 
bremsstrahlung is inconsistent with radio polarization measurements 
\citep{hk02}.  In fact, the detection of radio polarization indicates 
that the synchrotron mechanism is operating in the radio band.  In 
some cases, the broad-band (radio to X-ray) energy spectra can be 
fitted with a single power-law or a power-law with one break 
\citep{marshall01,hbw01,hardcastle02}, supporting a model where 
the broad-band jet spectrum is dominated by synchrotron emission.  
However, inverse Compton emission in the form of synchrotron
self-Compton (SSC) is known to dominate the X-ray emission for 
some AGN jets \citep{hk02}.

Other than XTE~J1550--564, the only X-ray binary where X-ray
jet spectra have been obtained is the well-known system SS~433
that resides in the radio nebula W50.  The large-scale
SS~433 X-ray jets extend up to $1^{\circ}$ from the X-ray 
binary, and the energy spectrum of the jet shows significant
spatial variation.  On scales of arcseconds and smaller, the
jets exhibit X-ray emission lines, indicating that thermal 
bremsstrahlung contributes to the X-ray emission 
\citep{marshall02,mfm02}.  However, on larger angular
scales ($15^{\prime}$ to $45^{\prime}$), emission lines are
not detected with {\em ASCA (Advanced Satellite for Cosmology 
and Astrophysics)} in the X-ray band \citep{yka94} and 
{\em RXTE} observations indicate that the spectrum extends to 
100~keV \citep{sp99}, indicating a non-thermal mechanism.

To study the implications of the various emission mechanisms 
for the XTE~J1550--564 jets, it is necessary to estimate the volume 
of the emitting region.  We assume that the shape of the emitting 
region is a section of a cone with its vertex at the compact object.  
The base of the conical section is at a distance of $d\tan\alpha_{b}$ 
from the compact object, where $\alpha_{b}$ is equal to the angular 
source separation given in Table~\ref{tab:results} plus 
$2^{\prime\prime}.5$.  The value of $2^{\prime\prime}.5$ comes from 
our estimate of the source extension obtained in \S3.2.  Although 
there is some uncertainty associated with the size of the extension 
for the eastern jet, we obtained a {\em Chandra} image of the western 
jet with much better statistics and measured a similar extension 
\citep{kaaret02}.  The top of the conical section is at a distance
of $d\tan\alpha_{t}$ from the compact object, where $\alpha_{t}$
is equal to the angular source separation minus $2^{\prime\prime}.5$.
As the opening angle of the cone is not well-constrained, we 
consider values for the full opening angle of $7.5^{\circ}$ 
(our upper limit for the eastern jet obtained in \S3.2) and
also $1^{\circ}$.  The latter may be more realistic as the upper 
limit on the opening angle for the western jet is $2^{\circ}$
\citep{kaaret02}.  In addition, we considered the distance 
range of $d = 2.8$-7.6~kpc in determining the volume.  A caveat is 
that the volume could be somewhat larger if $\theta$ (the 
angle between the jet axis and the line-of-sight) is small; 
however, the fact that the VLBI jets were two-sided and similar 
in brightness \citep{hannikainen01} makes this unlikely.  We
obtain a minimum volume of $3.8\times 10^{49}$~cm$^{3}$ for a 
distance of 2.8~kpc and an opening angle of $1^{\circ}$ and a 
maximum volume of $4.3\times 10^{52}$~cm$^{3}$ for $d = 7.6$~kpc 
and an opening angle of $7.5^{\circ}$.

The volume estimates allow for a calculation of the mass implied 
by the thermal bremsstrahlung fits to the energy spectrum (see 
\S3.3) for observation 1.  Assuming that the plasma consists of 
electrons and protons (i.e., $n_{e} = n_{i}$) and that $n_{e}$ is 
constant over the emitting volume gives 
$n_{e} = (6.5\times 10^{7})~d~\sqrt{N/V}$~cm$^{-3}$.  We obtain 
a minimum mass of $4\times 10^{28}$~g for $d = 2.8$~kpc, 
$V = 3.8\times 10^{49}$~cm$^{3}$ and $N = 4\times 10^{-5}$~cm$^{-5}$
and a maximum mass of $5\times 10^{30}$~g for $d = 7.6$~kpc, 
$V = 4.3\times 10^{52}$~cm$^{3}$ and $N = 1\times 10^{-4}$~cm$^{-5}$.
It is difficult to devise a scenario where the jet could contain
such large masses.  Even at the lower end of the mass range, it 
would take the black hole more than 1000 years to accumulate this 
much mass through accretion at a rate of $10^{18}$~g~s$^{-1}$, 
corresponding to the accretion rate for a luminosity of
$10^{38}$~erg~s$^{-1}$ and an accretion efficiency of 10\%.  It 
would also be difficult for the jet to start out with much less 
mass and then to entrain most of its mass from the ISM gradually
as it moves away from the compact object because momentum 
conservation would lead to rapid deceleration of the jet.  
However, we probably cannot rule out a scenario where the 
jet suddenly collides with a large amount of material that is 
far from the compact object such as material that is left from 
previous ejections.  Thermal bremsstrahlung may still be a possible 
X-ray emission mechanism if the jet has sufficient kinetic energy 
associated with its bulk motion to heat such material to X-ray 
temperatures.

Here, we assume that the X-ray and radio spectra can be described
by a single power-law with an index of $\alpha = -0.6$ as shown
in Figure~\ref{fig:broadband} to study the implications of a 
synchrotron X-ray emission mechanism.  A lower limit on the 
strength of the magnetic field in the jet, $B_{eq}$, comes 
from the assumption of equipartition between the magnetic and 
electron energy densities \citep{burbidge56}.  We derive the
equipartition magnetic field using
\begin{equation}
B_{eq} = \left(\frac{19 C_{12} L}{V}\right)^{2/7}~~~~~~~~~,
\end{equation}
\citep{pacholczyk70} where $L$ is the integrated luminosity 
and $C_{12}$ is an expression, given in \cite{pacholczyk70},
that depends on the lower and upper frequency limits and 
the spectral index ($\alpha$).  For both parameters, we used the 
frequency range from $1.384\times 10^{9}$ to $1.9\times 10^{18}$~Hz
and $\alpha = -0.6$.  We calculated $L$ using a flux density of
17~nJy at $7\times 10^{17}$~Hz (appropriate for observation 1)
and the range of distances and volumes derived in the previous
paragraphs.  Considering these distance and volume ranges, the 
lowest value of $B_{eq}$ we obtain is 134~$\mu$G for $d = 7.6$~kpc 
and $V = 4.3\times 10^{52}$~cm$^{3}$, and the highest value of 
$B_{eq}$ we obtain is 566~$\mu$G for $d = 2.8$~kpc and
$V = 3.8\times 10^{49}$~cm$^{3}$.  These values for $B_{eq}$
are somewhat higher than the values found from radio observations
of other Galactic objects (e.g., Rodriguez, Mirabel \& Marti 1992\nocite{rmm92}).
This is at least in part due to the fact that the equipartition
energy is higher because the spectrum for the XTE~J1550--564
jet extends to the X-ray band.

While it is important to keep in mind that the value for $B_{eq}$ 
assumes equipartition, we can use our result to obtain estimates of 
other physical jet quantities such as the Lorentz factor ($\gamma$) 
and lifetime ($t_{s}$) of the X-ray emitting electrons, the number
of electrons in the jet and the total mass of material in the jet.
From $\gamma(\nu)\sim (\nu/\nu_{L})^{1/2}$ \citep{bbr84}, where 
$\nu_{L}$ is the Larmor frequency, we obtain values of $\gamma$ 
at 1~keV between $1.2\times 10^{7}$ and $2.5\times 10^{7}$, and 
we obtain synchrotron lifetimes of 6-60~years at 1~keV using 
$t_{s} = 8\times 10^{8} B_{eq}^{-2} \gamma^{-1}$~s \citep{bbr84}, 
which is longer than the lifetime of the 
XTE~J1550--564 X-ray jet.  We note that if the flux decay
by a factor of 2-3 we observe over a period of about 4 months
during the {\em Chandra} observations is due to synchrotron
losses, this would imply a significantly shorter lifetime and 
probably a magnetic field strength above the equipartition value.
Following \cite{fender99}, the number of electrons in the jet
can be calculated from the parameters of the energy spectrum
($\alpha$ and luminosity) and $B_{eq}$, and we obtain a range
of values from $4\times 10^{44}$ to $1\times 10^{46}$ electrons.
Assuming one proton per electron gives values between 
$7\times 10^{20}$~g and $2\times 10^{22}$~g for the mass of
material in the jet.  For a mass accretion rate of $10^{18}$~g~s$^{-1}$, 
this amount of material could be accumulated in 700~s
and $2\times 10^{4}$~s for the two masses, respectively.  
These times fit with the likely ejection time as the large X-ray
and radio flare occurred 2 days before the jets were resolved 
using VLBI.

Although we have shown that a synchrotron mechanism is viable 
for the X-ray jet emission from XTE~J1550--564, we cannot 
immediately rule out inverse Compton or SSC.  However, we can
determine how much emission these mechanisms are likely to
contribute by comparing the photon energy density to the
magnetic energy density.  We calculate the magnetic energy
density assuming that only the radio emission has a synchrotron
origin, and we use the parameters from the power-law fits to
the radio emission ($\alpha = -0.82$ and a flux density of 
1.1 mJy at 5000~MHz) described in \S3.3.  Using the same distance
and volume ranges as above, we obtain $B_{eq} = 82$-344~$\mu$G
and a magnetic energy density ($B_{eq}^{2}/8\pi$) range of 
(3-50)$\times 10^{-10}$~erg~cm$^{-3}$.  We determine the
photon energy density ($u$) due to the source itself 
using $u = 3 L R/4 c V$ \citep{wys00}, where $R$ is the
linear size of the source, and we use the approximation 
$R = (3 V/4\pi)^{1/3}$.  It should be noted that $u$ is
distance independent, and the main uncertainty in its
determination is the opening angle of the cone that is
used to estimate $V$.  For opening angles of $1^{\circ}$
and $7.5^{\circ}$, respectively, $u$ is 
$6\times 10^{-12}$~erg~cm$^{-3}$ and $4\times 10^{-13}$~erg~cm$^{-3}$.
As these values are considerably less than the magnetic 
energy density, SSC should not be important.  Inverse
Compton from the interstellar radiation field (ISRF) or the 
cosmic microwave background (CMB) are also possibilities.
However, the maximum ISRF photon energy density in the
Galaxy is $2\times 10^{-11}$~erg~cm$^{-3}$ \citep{smr00}
and the CMB photon energy density is $3\times 10^{-13}$~erg~cm$^{-3}$, 
which are both considerably less than the magnetic energy
density.

\subsection{Jet Deceleration}

The mean proper motion for the eastern jet between its ejection
in 1998 September and the {\em Chandra} detections in 2000
of 32.9~mas day$^{-1}$ derived above is greater than the proper 
motion of $21.2\pm 7.2$~mas~day$^{-1}$ that we 
measure during 2000 June-September, indicating a reduction in 
the jet velocity.  The 1998 September VLBI observations of the 
radio jets provide information consistent with a higher initial 
ejection velocity. \cite{hannikainen01} find that the separation 
between the two radio jets increases at a rate of 115~mas~day$^{-1}$.  
Thus, with no additional information, we conclude that the initial 
proper motion for the eastern, approaching jet is 
at least 57.5~mas~day$^{-1}$, indicating that the apparent jet 
velocity decreased by a factor of at least $2.7\pm 0.9$.  This 
also implies a drop in the actual jet velocity; however, the size
of this drop depends on the distance to the source ($d$) and the 
angle between the jet axis and the line-of-sight ($\theta$), and 
these quantities are not well-constrained.  

Although the jet does decelerate, the fact that we detect a proper 
motion for the X-ray jet is in contrast to the apparently stationary 
radio lobes that are seen in GRS~1758--258 \citep{rmm92} and 
1E~1740.7--2942 \citep{mirabel92}.  However, the projected separation 
between XTE~J1550--564 and the X-ray jet (0.31-0.85~pc for 
$d = 2.8$-7.6~kpc) is a reasonable fraction of the compact object/radio 
lobe separations in the other two sources (2.5~pc in GRS~1758--258 and 
1.1~pc in 1E~1740.7--2942 assuming $d = 8.5$~kpc in both cases).  
Thus, the XTE~J1550--564 X-ray jet may represent an intermediate 
stage in the evolution of the jet as the bulk motion velocity decreases 
from relativistic velocities to zero velocity.

Observations of AGN suggest that their initially relativistic
jets also decelerate as they move away from the core of the
galaxy \citep{bridle94}.  A model has been developed to explain
jet deceleration where material in the interstellar medium 
(ISM) or intergalactic medium (IGM) is entrained into the jet 
\citep{bicknell94,blk96}.  As a consequence of momentum 
conservation, mass entrainment leads to deceleration of the jet.
It is not clear if the XTE~J1550--564 jet is entraining material,
but even if it is not, elastic collisions between the jet and
the ISM material will cause the jet to decelerate.  The level
of mass entrainment depends on how the jet pressure compares
to the ISM gas pressure.  The equipartition jet pressure 
($p_{eq} = B_{eq}^{2}/4\pi$) for the XTE~J1550--564 X-ray 
jet is between $10^{-9}$~erg~cm$^{-3}$ and a few times 
$10^{-8}$~erg~cm$^{-3}$ for the range of $B_{eq}$ values
derived above (134-566~$\mu$G assuming the X-rays have a
synchrotron origin).  These values for $p_{eq}$ are considerably
larger than the values typically measured for the ISM, which 
are between $10^{-13}$ and $10^{-12}$~erg~cm$^{-3}$ \citep{jt01},
and this may suggest that elastic collisions are more important
than entrainment in causing the jet to decelerate.  A related 
question is whether we are observing the motion of the bulk flow 
of material in the jet or the motion of a shock front.  
Although the latter is a possibility, further deceleration
of the eastern X-ray jet is reported in 
Kaaret et al.~(2002)\nocite{kaaret02}, indicating that a mechanism 
would be necessary to explain the deceleration of the shock.

It is also possible to derive the total internal energy for
the X-ray emitting electrons under the equipartition assumption.
For the range of source distances and jet opening angles used
above, we obtain energies between $5\times 10^{41}$~erg 
(for $d = 2.8$~kpc and an opening angle of $1^{\circ}$) and
$3\times 10^{43}$~erg (for $d = 7.6$~kpc and an opening
angle of $7.5^{\circ}$).  With the deceleration of the jet, 
it is interesting to consider whether the change in the bulk 
motion kinetic energy can account for the internal energy of 
the electrons.  From the {\em Chandra} proper motion measurement 
in 2000, the VLBI proper motion lower limit in 1998 and the total 
mass of protons in the jet of $7\times 10^{20}$~g derived above 
for an assumed distance of 2.8~kpc and an assumed opening angle 
of $1^{\circ}$, we obtain a lower limit on the change in
bulk motion kinetic energy ($\Delta$$KE$) of $2\times 10^{41}$~erg 
for $\theta = 43^{\circ}$.  For larger values of $\theta$, which 
are probably more likely for XTE~J1550--564, the lower limit on 
$\Delta$$KE$ increases gradually to $9\times 10^{41}$~erg, which 
is nearly a factor of 2 larger than the internal electron 
energy.  Although the internal electron energy estimates increase 
significantly with source distance and opening angle, for
$d > 3$~kpc, there are non-zero values for $\theta$ that
give initial jet velocities of $c$, leading to arbitrarily 
large values for $\Delta$$KE$.  We conclude that, for 
certain values of $\theta$, the bulk motion kinetic energy 
is sufficient to power the observed X-ray emission for most,
if not all, of the allowed distance range.  However, it is 
important to note that this calculation assumes that the total 
mass of protons in the jet is constant.  A significant level of 
entrainment from the ISM would invalidate this assumption.

\subsection{Observations of Quiescent Black Hole Transients}

The discovery of X-ray jets from XTE~J1550--564 has implications
for X-ray observations of black hole transients at low flux levels
or in quiescence using observatories without sufficient angular 
resolution to separate the black hole flux from the jet flux.
For example, if XTE~J1550--564 was observed with {\em ASCA} or 
{\em BeppoSAX (Satellite per Astronomia X)} during the times of 
our {\em Chandra} observations 2 and 3, the X-ray jet flux would 
significantly contaminate the black hole flux, but the X-ray jet 
would not be resolved due to the 1$^{\prime}$-3$^{\prime}$ angular 
resolution provided by these satellites.  For XTE~J1550--564, the 
measured X-ray jet fluxes are between $6\times 10^{-14}$ and 
$20\times 10^{-14}$ erg~cm$^{-2}$~s$^{-1}$ (0.3-8 keV), which is 
considerably brighter than 5 of the 6 black hole systems observed 
by {\em Chandra} in quiescence \citep{garcia01}.  For the 1996 March 
{\em ASCA} observation of the microquasar GRO~J1655--40 \citep{asai98}, 
it is conceivable that an X-ray jet was responsible for at least 
part of the detected flux.  The {\em ASCA} observation was made 
about 1.5 years after the discovery of powerful radio jets from 
this source \citep{tingay95,hr95}.  In addition, although it was 
thought that GRO~J1655--40 was in quiescence during the {\em ASCA} 
observation, the source flux was an order of magnitude lower 
during an observation made by {\em Chandra} in 2000 \citep{garcia01}.  
Finally, we note that if X-ray jet fluxes in other sources are 
similar to the level we observe for XTE~J1550--564, they will not 
significantly contaminate the non-imaging measurements made by 
{\em RXTE} as the quality of the PCA (Proportional Counter Array) 
background subtraction limits observations to sources brighter than 
a few$\times 10^{-12}$ erg~cm$^{-2}$~s$^{-1}$ (3-20 keV).

\section{Summary and Conclusions}

In this paper, we present an analysis of {\em Chandra} observations 
and an {\em ATCA} radio observation of XTE~J1550--564 made in 2000 
where an X-ray jet is detected to the east of the black hole in the 
X-ray and radio bands.  The discovery of the jet was first reported 
by \cite{corbel02b}.  The jet axis is aligned with the relativistic 
bipolar radio jets detected with VLBI in 1998 \citep{hannikainen01}.  
We argue that the material for the 2000 jet was likely ejected during 
the large X-ray and radio flare that occurred in 1998 \citep{wu02}, 
and that we detect the eastern jet (but not the western jet) during 
the 2000 observations because the eastern jet is the approaching jet.  
Although we do not detect a western jet for the 2000 observations, 
it is detected in the radio band \citep{corbel02a} and X-ray 
\citep{kaaret02} during observations made in 2002.  In fact, 
this paper is a result of a re-analysis of the 2000 {\em Chandra} 
data that was prompted by the radio detection of the western jet.

A main result of this work is the measurement of the proper
motion of the X-ray jet using three {\em Chandra} observations
made between 2000 June and 2000 September.  The X-ray jet 
moves away from XTE~J1550--564 with a proper motion of 
$21.2\pm 7.2$~mas~day$^{-1}$, and these observations represent
the first time that an X-ray jet proper motion measurement has been 
obtained for any accretion powered Galactic or extra-galactic 
source.  Comparing this value to the lower limit on the proper 
motion of the eastern jet from the 1998 VLBI observations of 
57.5~mas~day$^{-1}$ indicates that the jet has decelerated since 
its initial ejection.  Assuming the source distance is in the 
range from $d = 2.8$-7.6~kpc, the apparent jet velocity 
($\beta_{app}$) is between $0.34\pm 0.12$ and $0.93\pm 0.32$, 
and we obtain a lower limit on the actual X-ray jet velocity 
in 2000 of $\beta = 0.32\pm 0.10$ that is independent of the 
angle between the jet axis and the line-of-sight ($\theta$).  
There is evidence that the eastern jet is extended by 
$\pm 2^{\prime\prime}$-$3^{\prime\prime}$ in the direction of 
the proper motion.  The upper limit on the source extension in 
the perpendicular direction is $\pm 1^{\prime\prime}.5$, which 
corresponds to a jet opening angle of $<$$7.5^{\circ}$.

We consider thermal bremsstrahlung, synchrotron, SSC and inverse 
Compton mechanisms to explain the X-ray jet emission from 
XTE~J1550--564.  Assuming equipartition between the magnetic 
and electron energy densities, we derive a range of jet magnetic 
field strengths ($B_{eq}$) between 134~$\mu$G and 566~$\mu$G, 
depending on the source distance and jet opening angles between 
$1^{\circ}$ and $7.5^{\circ}$.  From $B_{eq}$, the range of Lorentz 
factors ($\gamma$) for the electrons responsible for the 1~keV 
X-rays is (1.2-2.7)$\times 10^{7}$, and the range of synchrotron 
lifetimes (at 1~keV) is 6-60~years.  The mass of material required 
by a synchrotron mechanism could be accumulated in the likely
jet ejection time of $<$2~days, while the mass required by
a thermal bremsstrahlung mechanism is about 8 orders of magnitude
higher, requiring mass from a very large number of ejection
events.  A comparison between the magnetic energy density
in the jet and estimates for the photon density suggest that
synchrotron emission should dominate over SSC and inverse
Compton mechanisms.  We conclude that a synchrotron mechanism 
is viable and appears to provide the simplest explanation for the 
observations.  The synchrotron hypothesis can be tested via 
constraints on the optical or IR flux from the jet or by 
obtaining better constraints on X-ray emission line strengths.

With a projected separation of 0.31-0.85~pc between 
XTE J1550--564 and the jet in 2000, the XTE~J1550--564 
jet is intermediate in size to the 0.02-0.06~pc relativistic 
radio jets that have been detected in several sources (e.g., 
GRS~1915+105 and GRO~J1655--40) and the stationary 1-3~pc 
radio lobes that have been seen in GRS~1758--258 and
1E~1740.7--2942.  Further, the fact that the XTE~J1550--564 
jets decelerated between 1998 and 2000 provides an additional 
connection between the two types of jets that have been 
observed previously.  It is likely that the deceleration occurs 
when the jet collides with the ISM.  However, the jet pressure
we derive from equipartition suggests that the pressure
is too high to entrain material, and that elastic collisions
of the jet with ISM material may be a better way to model
the jet evolution.  More work is necessary to understand
the details of the deceleration.  Finally, studying the
evolution of the XTE~J1550--564 jet underscores the fact that 
jets in Galactic X-ray binaries can provide useful information 
about jets in AGN.  In XTE~J1550--564, the time scale for 
material to travel the length of the jet is several years, 
while this process takes tens of thousands of years for a 
typical AGN.

\acknowledgements

JAT and SC acknowledge useful conversations with participants 
of the 4th Microquasar Workshop in Corsica, especially A. Celotti, 
S. Heinz and V. Dhawan.  JAT acknowledges useful conversations 
with G. Fossati, D. Fox, W. Heindl, R. Rothschild and G. Burbidge.  
We thank {\em Chandra} director H. Tananbaum for granting 
Director's Discretionary Time for the August and September 
observations described in this work.  RW was supported by NASA 
through Chandra Postdoctoral Fellowship grant number PF9-10010 
awarded by CXC, which is operated by SAO for NASA under contract 
NAS8-39073.  PK acknowledges partial support from NASA grant 
NAG5-7405 and Chandra grant G01-2034X.  JAT acknowledges partial 
support from NASA grant NAG5-10886.


\end{document}